# Density-dependent spin susceptibility and effective mass in monolayer MoSe$_2$

Chang Liu[1†], Tongtong Jia[1†], Zheng Sun[1†], Yu Gu[1], Fan Xu[1,2], Kenji Watanabe[3], Takashi Taniguchi[4], Jinfeng Jia[1,2,5,6], Shiyong Wang[1,2,5,6], Xiaoxue Liu[1,2,5,6], and Tingxin Li[1,2,5*]

[1]Key Laboratory of Artificial Structures and Quantum Control (Ministry of Education), School of Physics and Astronomy, Shanghai Jiao Tong University, Shanghai 200240, China

[2]Tsung-Dao Lee Institute, Shanghai Jiao Tong University, Shanghai, 201210, China

[3]Research Center for Electronic and Optical Materials, National Institute for Materials Science, 1-1 Namiki, Tsukuba 305-0044, Japan

[4]Research Center for Materials Nanoarchitectonics, National Institute for Materials Science, 1-1 Namiki, Tsukuba 305-0044, Japan

[5]Hefei National Laboratory, Hefei 230088, China

[6]Shanghai Research Center for Quantum Sciences, Shanghai 201315, China

†These authors contribute equally to this work.

*Emails: txli89@sjtu.edu.cn,

**Abstract:**

Atomically thin MoSe$_2$ is a promising platform for investigating quantum phenomena due to its large effective mass, high crystal quality, and strong spin-orbit coupling. In this work, we demonstrate a triple-gate device design with bismuth contacts, enabling reliable ohmic contact down to low electron densities, with a maximum Hall mobility of approximately 22,000 cm$^2$/Vs. Low-temperature transport measurements illustrate metal-insulator transitions, and density-dependent quantum oscillation sequences. Enhanced spin susceptibility and density-dependent effective mass are observed, attributed to interaction effects and valley polarization. These findings establish monolayer MoSe$_2$ as a versatile platform for further exploring interaction-driven quantum states.

**Main:**

Two-dimensional (2D) transition-metal dichalcogenide (TMDc) semiconductors, and their moiré superlattices, have attracted great interests from both the perspective of fundamental physics and device applications [1-3]. However, achieving reliable ohmic contact to atomically thin TMDc remains challenging [4,5], especially for hBN-encapsulated devices and at low temperatures and low carrier densities. Tremendous attempts have been made to improve the electrical contacts for TMDc semiconductors. For example, to reduce the Schottky barrier between the metal electrodes and 2D semiconductors, suitable work-functional metals have been used [6-18]; to reduce the Fermi-level pinning effects, tunneling contacts [19-21] and transferred-metal methods [22,23] have been applied. Moreover, phase engineering [24-26] or inducing heavily doping [27-29] in contact regions have also been widely used to achieve ohmic contact for TMDc semiconductors. Recent experiments report [30-32] that the semimetal bismuth (Bi) and antimony (Sb) form excellent *n*-type ohmic contacts with monolayer MoS$_2$, approaching the theoretical limit for contact resistance. Overall, at cryogenic temperatures, reasonably good electrical contacts to atomically thin TMDc semiconductors has been successfully achieved mainly in *p*-type WSe$_2$ [6-8,17,18,28,29], *p*-type MoTe$_2$ [13-16], and *n*-type MoS$_2$ [6,9,10,33]. This is crucial for the recent observations of integer and fractional quantum anomalous Hall effects, quantum spin Hall effects and superconductivities in WSe$_2$/MoTe$_2$ [13], twisted MoTe$_2$ [14-16] and twisted WSe$_2$ [17,18] moiré systems. It has also enabled the observation of fractional quantum Hall effects under high magnetic fields in monolayer WSe$_2$ [28] and bilayer MoS$_2$ [32].

Among various TMDc semiconductors, MoSe$_2$ has attracted significant attentions due to its large effective mass $m^*$ and high crystal quality [34]. Recent optical and scanning tunneling spectroscopy studies have reported signatures of Wigner crystals in monolayer and bilayer MoSe$_2$ [35,36]. Additionally, thermodynamic and transport evidence of interlayer exciton insulators have been observed in MoSe$_2$/WSe$_2$ heterostructure devices [37-39]. Reliable electrical transport measurements with ohmic contacts at low temperatures and low carrier densities are highly desirable to provide smoking-gun evidences of Wigner crystals and exciton superfluidity. Previously, only one study [11] demonstrated that palladium (Pd) can form reasonably good ohmic contacts with atomically thin MoSe$_2$ at low temperatures, enabling quantum transport measurements of monolayer and bilayer MoSe$_2$ devices, but only at relatively high electron densities ($> 2\times10^{12}$ cm$^{-2}$). Here we report a new device structure with Bi contacts, enabling reliable electrical contact to high quality monolayer MoSe$_2$ devices at electron density as low as $0.3\times10^{12}$ cm$^{-2}$. The device achieves a maximum Hall mobility of approximately 22,000 cm$^2$/Vs. Low-temperature transport measurements reveal density-dependent quantum oscillation sequences, which can be attributed to interaction-enhanced spin susceptibility. Additionally, we also observe a density-dependent effective mass, providing further insight into the band structure and electronic properties of monolayer MoSe$_2$.

Similar to the device structure used in our previous studies on twisted bilayer MoTe$_2$ [15,40], we employed a triple-gate design for monolayer MoSe$_2$ devices. In this device configuration (as shown in Fig. 1), the top gate and bottom gate are used to independently tune the electric field and carrier density of the sample, while a silicon gate is used to achieve high electron densities in the contact region, enabling reliable electrical contacts. Unlike in twisted MoTe$_2$ devices, where the 2D van der Waals metal TaSe$_2$ with high work function were used to form *p*-type contacts for MoTe$_2$, we choose the low work function semimetal Bi [30] as the contact material for *n*-type MoSe$_2$. Figure 1a illustrates a schematic representation of the sample fabrication and assembly processes. Bulk crystals of graphite, hexagonal boron nitride (h-BN), and MoSe$_2$ (commercial crystals from HQ graphene, grown by chemical vapor transport methods) were exfoliated onto a Si/SiO$_2$ (285 nm) substrate. Graphite flakes of 2-4 nm thickness and h-BN flakes of 20-30 nm thickness were used as the top gate (TG) and bottom gate (BG). First of all, top graphite and top h-BN layers were assembled using dry transfer methods [41], and released on to a Si/SiO$_2$ substrate with markers. Subsequently, electron beam lithography (EBL) and reactive ion etching (RIE) were employed to etch contact windows through the top h-BN layer. Monolayer MoSe$_2$ were then picked up by the etched top graphite/h-BN stack, followed by sequential pick-up of the bottom h-BN and bottom graphite layers. Then the whole stack was released on a Si/SiO$_2$ substrate again. The shape of the contact electrodes is precisely defined using EBL, ensuring an accurate alignment with the pre-patterned windows on the top h-BN layer. In order to form ohmic contacts, Bi/Au (15nm/35nm) films were deposited on the monolayer MoSe$_2$ through pre-patterned windows using an electron beam evaporator with a vacuum better than $5\times10^{-4}$ Pa. Finally, EBL and RIE processes were employed again to define the Hall bar shape of the device. Figure 1b and 1c show the optical microscope image and the schematic device structure, respectively.

Figure 2a shows the two-terminal resistance $R_2$, four-terminal longitudinal resistance $R_{xx}$, and contact resistance $R_c$ as a function of the electron density $n$ (controlled by electrostatic gating) of a typical monolayer MoSe$_2$ device at temperature $T=1.8$ K. The contact resistance $R_c$ is estimated by $2R_c = R_2 - \frac{L}{W}\rho_{xx}$, where $L$ is the length between the source and drain, $W$ is the width of the Hall bar, and $\rho_{xx}$ can be estimated by measured $R_{xx}$ and the Hall bar geometry. It can be seen that the $R_c$ remains below 20 kΩ down to an electron density of approximately $0.4\times10^{12}$ cm$^{-2}$ at $T=1.8$ K. Fig. 2b shows the Hall mobility $\mu_H$ as a function of $n$, which reaches a maximum value of approximately 22,000 cm$^2$/Vs at $n$ about $2.5\times10^{12}$ cm$^{-2}$. Figure 2c displays the $R_{xx}$ as a function of $T$ at $n$ between $0.22\times10^{12}$ cm$^{-2}$ and $0.69\times10^{12}$ cm$^{-2}$, where the critical density for the metal-insulator transition is identified at $n$ of approximately $0.37\times10^{12}$ cm$^{-2}$. Figure 2d and 2e illustrate the measured $R_{xx}$ as a function of perpendicular magnetic field ($B$) in monolayer MoSe$_2$ at $T$ =1.8 K, at two distinct electron densities. Well-resolved Shubnikov-de Haas (SdH) oscillations can be observed for $B$ > 3-4 T, corresponding to a quantum mobility $\mu_Q$ around 3,000 cm$^2$/Vs. Both the $\mu_H$ and $\mu_Q$ are the highest values reported to date in monolayer MoSe$_2$.

Figure 3a further illustrates the 2D map of $R_{xx}$ as a function of $B$ and $n$. Notably, we observe that the degeneracy of SdH oscillations varies with $n$. As shown in Fig. 3b (a 10 T linecut from Fig. 3a), for $n$ < ~1.5×10$^{12}$ cm$^{-2}$, the SdH oscillations are single-fold degenerate, with Landau level (LL) filling factor $\nu_{LL}$ up to 6. At higher carrier densities, the quantum oscillations are predominantly 2-fold degenerate, although the $\nu_{LL}$ sequences switch between odd- and even-integer dominance. When $n$ exceeds approximately 12.5×10$^{12}$ cm$^{-2}$ (indicated by the vertical dashed line in Fig. 3a), the quantum oscillations become four-fold degenerate. These observations can be well understood in terms of the MoSe$_2$ band structure and the density-dependent, interaction-enhanced Zeeman splitting. Figure 3c shows the schematic evolution of the conduction band alignment under finite $B$ at the $K$ and $K'$ valley of monolayer MoSe$_2$, as a function of $n$. On one hand, due to the Ising spin-orbit coupling, the spin degeneracy is lifted, thus the low energy band is two-fold degenerated and spin-valley locked. At higher carrier densities, both spin subbands are populated in each valley, resulting in four-fold degeneracy (Regime III in Fig. 3c). These can qualitatively explain the four-fold degenerated quantum oscillations observed at $n$ > ~12.5×10$^{12}$ cm$^{-2}$ and two-fold degenerated quantum oscillations observed at ~12.5×10$^{12}$ cm$^{-2}$ > $n$ > ~1.5×10$^{12}$ cm$^{-2}$ in Fig. 2b. On the other hand, the Zeeman splitting $E_Z = g^*\mu_B B$ ($g^*$ is the effective $g$-factor and $\mu_B = e\hbar/2m_0$ is the Bohr magneton, with $\hbar$ being the reduced Planck constant and $m_0$ the free electron mass) is related to the interaction strength of electrons, making it density-dependent, as previously reported in atomically thin WSe$_2$ [7,8,42,43], MoS$_2$ [9,10], MoSe$_2$ [11], as well as in GaAs and AlAs 2D electron gas systems [44,45]. Consequently, under an appropriate range of $B$ and $n$, the conduction band of monolayer MoSe$_2$ become valley and spin polarized (Regime I in Fig. 3c), leading to the observed single-fold degenerate quantum oscillations.

The ability to measure transport properties at low temperature and high magnetic fields, down to low electron densities in monolayer MoSe2 device, enables the unambiguous determination of the ratio between $E_Z$ and the cyclotron energy $E_c = \hbar\omega_c = \hbar eB/m^*$. The ratio, $E_Z/E_c = g^*m^*/2m_0$, is directly proportional to the spin susceptibility $\chi^* = g^*m^*\mu_B^2/2\pi\hbar^2$. We can infer the value of $E_Z/E_c$ in several ways. Firstly, from the $R_{xx}$-$n$-$B$ map, we could identify certain features at specific $n$ and $B$ points (marker by arrows in Fig. 3a), where the Landau fan degeneracy transits from one to two. Namely, at these particular $n$-values, the $E_Z/E_c$ ratios can be estimated by counting the number of polarized LLs. This allows us to determine the $n$ values corresponding to $E_Z/E_c$= 8, 7, 6, respectively. Secondly, the quantum oscillation sequences switch between odd- and even-integer dominance as a function of $n$, depending on the specific value of $E_Z/E_c$. As illustrated in the schematic of Fig. 3b, the even (odd) values of $E_Z/E_c$ lead to quantum Hall states with even (odd) $v_{LL}$, and in between of the even-integer and odd-integer dominant quantum Hall states, the $E_Z/E_c$ values should be closer to half-integer values. Since the ratio of $E_Z/E_c$ can only change continuously in intervals of 1, we could then infer the $E_Z/E_c$ values to be 5.5, 4.5 and 3.5 for the even-odd transition regions (shaded by blue in Fig. 3b). Finaly, at low electron densities, valley polarization condition can be easily achieved under relatively small $B$, even before the onset of SdH oscillations, resulting in a pronounced positive magnetoresistance until the $E_Z$ is equal to the Fermi energy, as previously reported [11,45]. The critical magnetic field $B_p$ for achieving valley polarization can be roughly determined from the magnetoresistance data, as illustrated in Fig. 3d. Importantly, $B_p$ is related to the $E_Z/E_c$ ratio, as given by $B_p = 2hn/(eg^*m^*/m_0) = hn/(eE_Z/E_c)$.

Fig. 3e summarize the obtained $E_Z/E_c$ values as a function of $n$. The enhancement of $E_Z/E_c$ (or $\chi^*$) is proportional to $n^{-1/2}$ with decreasing $n$, as illustrated by the dashed line in Fig. 3e. In two-dimensional electronic system with parabolic band dispersions, the Coulomb interaction energy decays slower than the kinetic energy, thus become more dominant at low carrier densities, characterized by the Wigner-Seitz radius $r_s = e^2 m^*/(4\pi\varepsilon_0\varepsilon\hbar^2\sqrt{\pi n}) \propto n^{-1/2}$. Here $\varepsilon_0$ and $\varepsilon$ denote the vacuum permittivity and dielectric constant, respectively. Therefore, the observed enhancement in spin susceptibility, following an $n^{-1/2}$ dependence, is likely attributable to the strengthened interaction effects, as previously reported [7-11, 42-45].

Interestingly, we found the $m^*$ of monolayer MoSe$_2$ is also *n* dependent. Figure 4a and 4b show the temperature dependence of the amplitude of quantum oscillations at *n* =1.2×10$^{12}$ cm$^{-2}$ and *n* =3.7×10$^{12}$ cm$^{-2}$, respectively. The $m^*$ can be deduced from the temperature dependent data based on the Lifshitz-Kosevitch formula, $R_T = \frac{2\pi^2 k_B T/\hbar\omega_c}{\sinh(2\pi^2 k_B T/\hbar\omega_c)}$. As a result, we found that the $m^*$ value at *n* =1.2×10$^{12}$ cm$^{-2}$ and *n* =3.7×10$^{12}$ cm$^{-2}$ is approximately $0.5\,m_0$ and $0.85\,m_0$, respectively, and with no significant dependence on *B* at each electron density (Fig. 4c). Notably, the $m^*$ value at lower electron density is closer to calculated $m^*$ values for monolayer MoSe$_2$ [46-48], and the $m^*$ value at higher electron density is consistent with previous experiments [11]. Figure 4d further illustrates $m^*$ as a function of *n*, deduced based on the temperature dependent quantum oscillations under *B* around 10 T. Clearly, the $m^*$ exhibits a sudden drop at low electron densities. The orange-shaded region in Fig. 4d highlights the valley polarization density range at *B* = 10 T, estimated based on the density-dependent $E_z/E_c$ shown in Fig. 3e. It demonstrates that the reduction in $m^*$ is closely related to the valley polarization. The underlying mechanism for such density-dependent $m^*$ needs further experimental and theoretical studies.

Such density-dependent $m^*$ has important implications for understanding electronic properties and band structures of monolayer MoSe$_2$. For example, theories predict [49] that Wigner crystals to be the ground state of a clean two-dimensional electronic system at low carrier densities with $r_s > 30$. For monolayer MoSe$_2$, it corresponds to $n \sim 0.13 \times 10^{12}$ cm$^{-2}$ when using $m^* = 0.5\, m_0$ and $n \sim 0.36 \times 10^{12}$ cm$^{-2}$ when using $m^* = 0.85 m_0$, assuming a dielectric constant $\varepsilon$ to be ~5 for typical h-BN encapsulated devices [35]. In addition, in Fig. 3a, we show that the second spin-split band starts to be populated at $n \sim 12.5 \times 10^{12}$ cm$^{-2}$, taking the $m^* = 0.85 m_0$ into account, we can estimate the spin splitting energy in MoSe$_2$ conduction band is around 35 meV.

In conclusion, our work establishes monolayer MoSe$_2$ as an exceptional platform for quantum transport studies by achieving reliable ohmic contacts at low carrier densities. The observation of density-dependent spin susceptibility and effective mass, extends the understanding of the interplay between Zeeman splitting, valley polarization, and interaction effects in monolayer MoSe$_2$. Our work paves the way for future explorations of novel quantum phases based on atomically thin MoSe$_2$ and their moiré structures.

**Reference**

1. Mak, K. F.; Shan, J. Semiconductor Moiré Materials. Nat. Nanotechnol. 2022, 17 (7), 686-695.

2. Cao, W.; Bu, H.; Vinet, M.; Cao, M.; Takagi, S.; Hwang, S.; Ghani, T.; Banerjee, K. The Future Transistors. Nature 2023, 620 (7974), 501-515.

3. Kim, K. S.; et al. The future of two-dimensional semiconductors beyond Moore's law. Nat. Nanotechnol. 2024, 19, 895–906.

4. Allain, A.; Kang, J.; Banerjee, K.; Kis, A. Electrical Contacts to Two-Dimensional Semiconductors. Nat. Mater. 2015, 14 (12), 1195-1205.

5. Wang, Y.; Chhowalla, M. Making Clean Electrical Contacts on 2D Transition Metal Dichalcogenides. Nat. Rev. Phys. 2022, 4 (2), 101-112.

6. Xu, S.; Wu, Z.; Lu, H.; Han, Y.; Long, G.; Chen, X.; Han, T.; Ye, W.; Wu, Y.; Lin, J.; et al. Universal Low-Temperature Ohmic Contacts for Quantum Transport in Transition Metal Dichalcogenides. 2D Mater. 2016, 3 (2), 021007.

7. Movva, H. C. P.; Fallahazad, B.; Kim, K.; Larentis, S.; Taniguchi, T.; Watanabe, K.; Banerjee, S. K.; Tutuc, E. Density-Dependent Quantum Hall States and Zeeman Splitting in Monolayer and Bilayer WSe2. Phys. Rev. Lett. 2017, 118 (24), 247701.

8. Xu, S.; Shen, J.; Long, G.; Wu, Z.; Bao, Z.-q.; Liu, C.-C.; Xiao, X.; Han, T.; Lin, J.; Wu, Y.; et al. Odd-Integer Quantum Hall States and Giant Spin Susceptibility in p-Type Few-Layer WSe2. Phys. Rev. Lett. 2017, 118 (6), 067702.

9. Pisoni, R.; Kormányos, A.; Brooks, M.; Lei, Z.; Back, P.; Eich, M.; Overweg, H.; Lee, Y.; Rickhaus, P.; Watanabe, K.; et al. Interactions and Magnetotransport through Spin-Valley Coupled Landau Levels in Monolayer MoS2. Phys. Rev. Lett. 2018, 121 (24), 247701.

10. Lin, J.; Han, T.; Piot, B. A.; Wu, Z.; Xu, S.; Long, G.; An, L.; Cheung, P.; Zheng, P. P.; Plochocka, P.; et al. Determining Interaction Enhanced Valley Susceptibility in Spin-Valley-Locked MoS2. Nano Lett. 2019, 19 (3), 1736-1742.

11. Larentis, S.; Movva, H. C. P.; Fallahazad, B.; Kim, K.; Behroozi, A.; Taniguchi, T.; Watanabe, K.; Banerjee, S. K.; Tutuc, E. Large Effective Mass and Interaction-Enhanced Zeeman Splitting of K-Valley Electrons in MoSe2. Phys. Rev. B 2018, 97 (20), 201407.

12. Wang, Y.; Kim, J. C.; Wu, R. J.; Martinez, J.; Song, X.; Yang, J.; Zhao, F.; Mkhoyan, A.; Jeong, H. Y.; Chhowalla, M. Van der Waals Contacts between Three-Dimensional Metals and Two-Dimensional Semiconductors. Nature 2019, 568 (7750), 70-74.

13. Li, T.; Jiang, S.; Shen, B.; Zhang, Y.; Li, L.; Tao, Z.; Devakul, T.; Watanabe, K.; Taniguchi, T.; Fu, L.; et al. Quantum Anomalous Hall Effect from Intertwined Moiré Bands. Nature 2021, 600 (7890), 641-646.

14. Park, H.; Cai, J.; Anderson, E.; Zhang, Y.; Zhu, J.; Liu, X.; Wang, C.; Holtzmann, W.; Hu, C.; Liu, Z.; et al. Observation of Fractionally Quantized Anomalous Hall Effect. Nature 2023, 622 (7981), 74-79.

15. Xu, F.; Sun, Z.; Jia, T.; Liu, C.; Xu, C.; Li, C.; Gu, Y.; Watanabe, K.; Taniguchi, T.; Tong, B.; et al. Observation of Integer and Fractional Quantum Anomalous Hall Effects in Twisted Bilayer MoTe2. Phys. Rev. X 2023, 13 (3), 031037.

16. Kang, K.; Shen, B.; Qiu, Y.; Zeng, Y.; Xia, Z.; Watanabe, K.; Taniguchi, T.; Shan,

J.; Mak, K. F. Evidence of the Fractional Quantum Spin Hall Effect in Moiré MoTe2. Nature 2024, 628 (8008), 522-526.

17. Xia, Y.; Han, Z.; Watanabe, K.; et al. Superconductivity in twisted bilayer WSe2. Nature 2024, 635 (8039).

18. Guo, Y.; Pack, J.; Swann, J.; Holtzman, L.; Cothrine, M.; Watanabe, K.; Taniguchi, T.; Mandrus, D.; Barmak, K.; Hone, J.; Millis, A. J.; Pasupathy, A. N.; Dean, C. R. Superconductivity in twisted bilayer $WSe_2$. arXiv preprint 2024, arXiv:2406.03418.

19. Wang, J.; Yao, Q.; Huang, C.-W.; Zou, X.; Liao, L.; Chen, S.; Fan, Z.; Zhang, K.; Wu, W.; Xiao, X.; et al. High Mobility MoS2 Transistor with Low Schottky Barrier Contact by Using Atomic Thick h-BN as a Tunneling Layer. Adv. Mater. 2016, 28 (37), 8302-8308.

20. Cui, X.; Shih, E.-M.; Jauregui, L. A.; Chae, S. H.; Kim, Y. D.; Li, B.; Seo, D.; Pistunova, K.; Yin, J.; Park, J.-H.; et al. Low-Temperature Ohmic Contact to Monolayer MoS2 by van der Waals Bonded Co/h-BN Electrodes. Nano Lett. 2017, 17 (8), 4781-4786.

21. Ghiasi, T. S.; Quereda, J.; van Wees, B. J. Bilayer h-BN Barriers for Tunneling Contacts in Fully-Encapsulated Monolayer MoSe2 Field-Effect Transistors. 2D Mater. 2019, 6 (1), 015002.

22. Liu, Y.; Guo, J.; Zhu, E.; Liao, L.; Lee, S.-J.; Ding, M.; Shakir, I.; Gambin, V.; Huang, Y.; Duan, X. Approaching the Schottky–Mott Limit in Van der Waals Metal–Semiconductor Junctions. Nature 2018, 557 (7707), 696-700.

23. Telford, E. J.; Benyamini, A.; Rhodes, D.; Wang, D.; Jung, Y.; Zangiabadi, A.; Watanabe, K.; Taniguchi, T.; Jia, S.; Barmak, K.; Pasupathy, A. N.; Dean, C. R.; Hone, J. Via Method for Lithography Free Contact and Preservation of 2D Materials. Nano Lett. 2018, 18, (2), 1416–1420.

24. Kappera, R.; Voiry, D.; Yalcin, S. E.; Branch, B.; Gupta, G.; Mohite, A. D.; Chhowalla, M. Phase-Engineered Low-Resistance Contacts for Ultrathin MoS2 Transistors. Nat. Mater. 2014, 13 (12), 1128-1134.

25. Cho, S.; Kim, S.; Kim, J. H.; Zhao, J.; Seok, J.; Keum, D. H.; Baik, J.; Choe, D.-H.; Chang, K. J.; Suenaga, K.; et al. Phase Patterning for Ohmic Homojunction Contact in MoTe2. Science 2015, 349 (6248), 625-628.

26. Cai, X.; Wu, Z.; Han, X.; Chen, Y.; Xu, S.; Lin, J.; Han, T.; He, P.; Feng, X.; An, L.; et al. Bridging the Gap between Atomically Thin Semiconductors and Metal Leads. Nat. Commun. 2022, 13 (1), 1777.

27. Wang, J. I. J.; Yang, Y.; Chen, Y.-A.; Watanabe, K.; Taniguchi, T.; Churchill, H. O. H.; Jarillo-Herrero, P. Electronic Transport of Encapsulated Graphene and WSe2 Devices Fabricated by Pick-up of Prepatterned hBN. Nano Lett. 2015, 15 (3), 1898-1903.

28. Pack, J.; Guo, Y.; Liu, Z.; Jessen, B. S.; Holtzman, L.; Liu, S.; Cothrine, M.; Watanabe, K.; Taniguchi, T.; Mandrus, D. G.; et al. Charge-Transfer Contacts for the Measurement of Correlated States in High-Mobility WSe2. Nat. Nanotechnol. 2024, 19 (7), 948-954.

29. Xie, J.; Zhang, Z.; Zhang, H.; Nagarajan, V.; Zhao, W.; Kim, H.-L.; Sanborn, C.; Qi, R.; Chen, S.; Kahn, S.; Watanabe, K.; Taniguchi, T.; Zettl, A.; Crommie, M. F.; Analytis, J.; Wang, F. Low Resistance Contact to P-Type Monolayer WSe2. Nano Lett. 2024, 24 (20), 5937–5943.

30. Shen, P.-C.; Su, C.; Lin, Y.; Chou, A.-S.; Cheng, C.-C.; Park, J.-H.; Chiu, M.-H.; Lu, A.-Y.; Tang, H.-L.; Tavakoli, M. M.; et al. Ultralow Contact Resistance between Semimetal and Monolayer Semiconductors. Nature 2021, 593 (7858), 211-217.

31. Li, W.; Gong, X.; Yu, Z.; Ma, L.; Sun, W.; Gao, S.; Köroğlu, Ç.; Wang, W.; Liu, L.; Li, T.; et al. Approaching the Quantum Limit in Two-Dimensional Semiconductor Contacts. Nature 2023, 613 (7943), 274-279.

32. Zhao, S.; Huang, J.; Crépel, V.; Xiong, Z.; Wu, X.; Zhang, T.; Wang, H.; Han, X.; Li, Z.; Xi, C.; et al. Fractional Quantum Hall Phases in High-Mobility n-Type Molybdenum Disulfide Transistors. Nat. Electron. 2024. https://doi.org/10.1038/s41928-024-01274-1

33. Cui, X.; Lee, G.-H.; Kim, Y. D.; Arefe, G.; Huang, P. Y.; Lee, C.-H.; Chenet, D. A.; Zhang, X.; Wang, L.; Ye, F.; et al. Multi-Terminal Transport Measurements of MoS2 Using a Van der Waals Heterostructure Device Platform. Nat. Nanotechnol. 2015, 10 (6), 534-540.

34. Liu, S.; Liu, Y.; Holtzman, L.; Li, B.; Holbrook, M.; Pack, J.; Taniguchi, T.; Watanabe, K.; Dean, C. R.; Pasupathy, A. N.; Barmak, K.; Rhodes, D. A.; Hone, J. Two-Step Flux Synthesis of Ultrapure Transition-Metal Dichalcogenides. ACS

Nano 2023, 17, 16587–16596.

35. Smoleński, T.; Dolgirev, P. E.; Kuhlenkamp, C.; Popert, A.; Shimazaki, Y.; Back, P.; Lu, X.; Kroner, M.; Watanabe, K.; Taniguchi, T.; et al. Signatures of Wigner Crystal of Electrons in a Monolayer Semiconductor. Nature 2021, 595 (7865), 53-57.

36. Xiang, Z.; Li, H.; Xiao, J.; Naik, M. H.; Ge, Z.; He, Z.; Chen, S.; Nie, J.; Li, S.; Jiang, Y.; et al. Quantum Melting of a Disordered Wigner Solid. arXiv preprint 2024, arXiv:2402.05456

37. Ma, L.; Nguyen, P.X.; Wang, Z.; et al. Strongly correlated excitonic insulator in atomic double layers. Nature 2021, 598 (7882), 585–589.

38. Nguyen, P. X.; Ma, L.; Chaturvedi, R.; Watanabe, K.; Taniguchi, T.; Shan, J.; Mak, K. F. Perfect Coulomb drag in a Dipolar Excitonic Insulator. arXiv preprint 2023, arXiv:2309.14940

39. Qi, R.; Joe, A. Y.; Zhang, Z.; Xie, J.; Feng, Q.; Lu, Z.; Wang, Z.; Taniguchi, T.; Watanabe, K.; Tongay, S.; et al. Perfect Coulomb Drag and Exciton Transport in an Excitonic Insulator. arXiv preprint 2023, arXiv:2309.15357

40. F. Xu.; et al. Interplay between topology and correlations in the second moiré band of twisted bilayer MoTe2. arXiv preprint 2024, arXiv: 2406.09687

41. L. Wang.; et al. One-Dimensional Electrical Contact to a Two-Dimensional Material. Science 2013, 342 (6158) ,614-617.

42. Gustafsson, M. V.; Yankowitz, M.; Forsythe, C.; Rhodes, D.; Watanabe, K.; Taniguchi, T.; Hone, J.; Zhu, X.; Dean, C. R. Ambipolar Landau Levels and Strong Band-Selective Carrier Interactions in Monolayer WSe2. Nat. Mater. 2018, 17 (5), 411-415.

43. Shi, Q.; Shih, E.-M.; Gustafsson, M. V.; Rhodes, D. A.; Kim, B.; Watanabe, K.; Taniguchi, T.; Papić, Z.; Hone, J.; Dean, C. R. Odd- and Even-Denominator Fractional Quantum Hall States in Monolayer WSe2. Nat. Nanotechnol. 2020, 15 (7), 569-573.

44. Zhu, J.; Stormer, H. L.; Pfeiffer, L. N.; Baldwin, K. W.; West, K. W. Spin Susceptibility of an Ultra-Low-Density Two-Dimensional Electron System. Phys. Rev. Lett. 2003, 90 (5), 056805.

45. Vakili, K.; Shkolnikov, Y. P.; Tutuc, E.; De Poortere, E. P.; Shayegan, M. Spin Susceptibility of Two-Dimensional Electrons in Narrow AlAs Quantum Wells. Phys. Rev. Lett. 2004, 92 (22), 226401.

46. A. Kormányos.; V. Zólyomi.; N. D. Drummond.; and G. Burkard.; Spin-Orbit Coupling, Quantum Dots, and Qubits in Monolayer Transition Metal Dichalcogenides. Phys. Rev. X. 2014, 4 (1), 011034.

47. A. Kormányos.; G. Burkard, M. Gmitra.; J. Fabian, V. Zólyomi.; N. D. Drummond.; and V. Fa`lko.; k·p theory for two-dimensional transition metal dichalcogenide semiconductors. 2D Mater. 2015 2 (2), 022001.

48. D. Wickramaratne.; F. Zahid.; and R. K. Lake.; Electronic and thermoelectric properties of few-layer transition metal dichalcogenides. The Journal of Chemical Physics. 2014 140 (12), 124710.

49. Drummond, N. D.; and Needs, R. J.; Phase diagram of the low-density two-dimensional homogeneous electron gas. Phys. Rev. Lett. 2009 102, 126402.

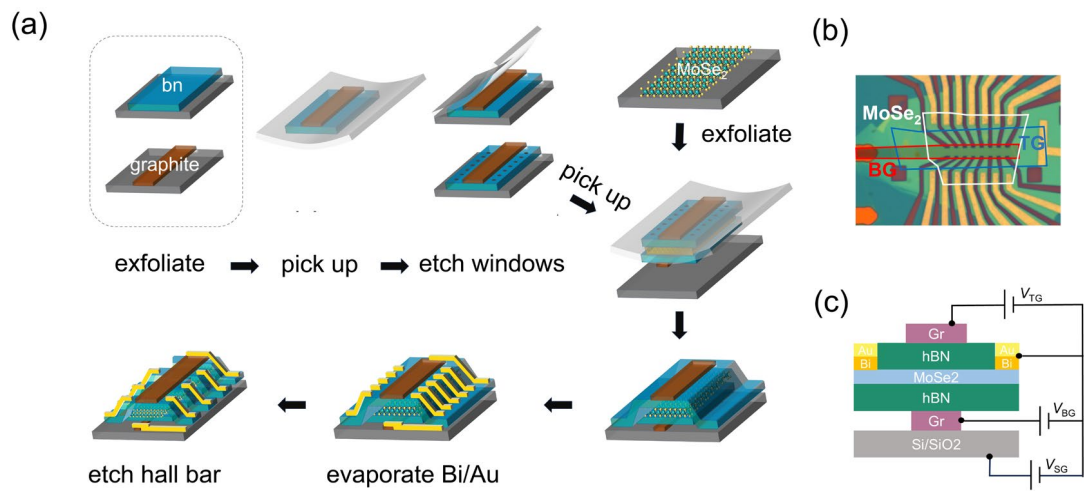

**Figure 1.** (a) Schematic diagrams illustrating the fabrication processes for triple-gated monolayer MoSe$_2$ device. (b) Optical micrograph of the device. The shapes and positions of the bottom gate (BG), top gate (TG), and monolayer MoSe$_2$ flakes are highlighted by lines with different colors. (c) Side view of the device structure. The carrier density within the Hall-bar channel is tuned by the top and bottom graphite gates, while the contact regions are heavily doped via the global Si/SiO$_2$ gate.

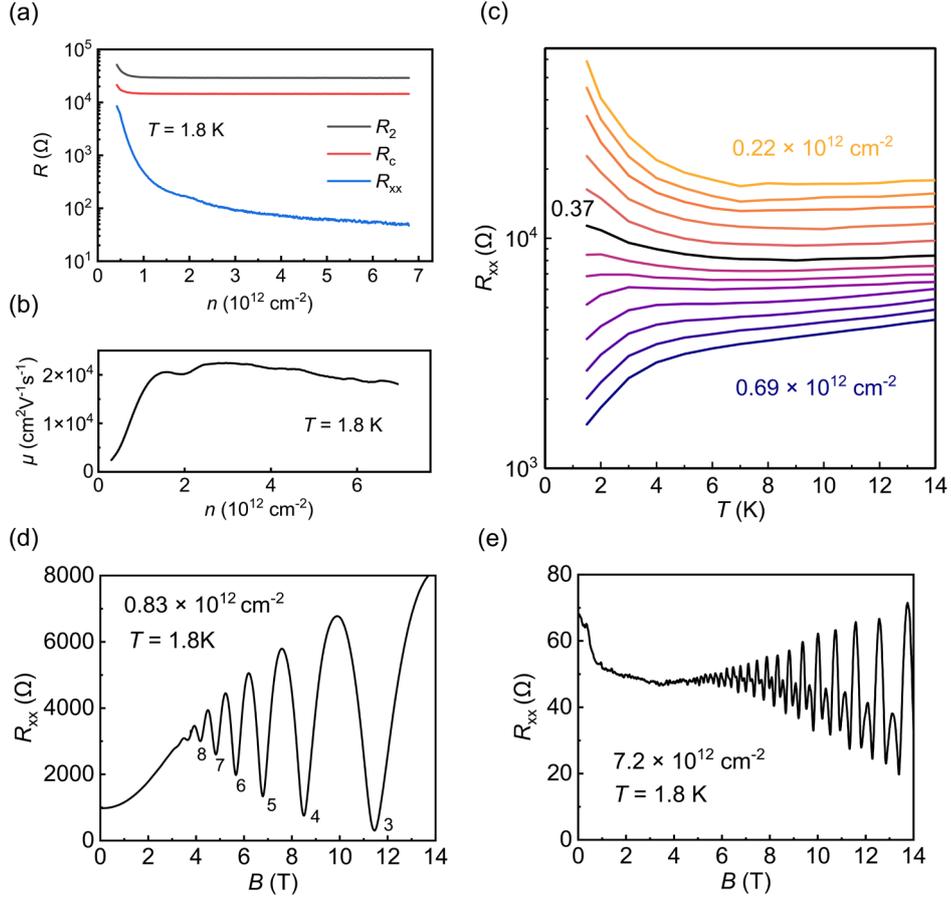

**Figure 2.** (a) Two-terminal resistance $R_2$, four-terminal longitudinal resistance $R_{xx}$, and contact resistance $R_c$ as a function of electron density at 1.8 K. (b) Hall mobility as a function of electron density at 1.8 K. (c) Measured $R_{xx}$ as a function of $T$, for $n$ from $0.22\times10^{12}$ cm$^{-2}$ to $0.69\times10^{12}$ cm$^{-2}$, revealing a metal-insulator transition occurring near $0.37\times10^{12}$ cm$^{-2}$. (d) $R_{xx}$ as a function of perpendicular magnetic field $B$ at 1.8 K at $n = 0.83\times10^{12}$ cm$^{-2}$ (d) and $n = 7.2\times10^{12}$ cm$^{-2}$ (e). The Landau level filling factors $\nu_{LL}$ is marked in (d).

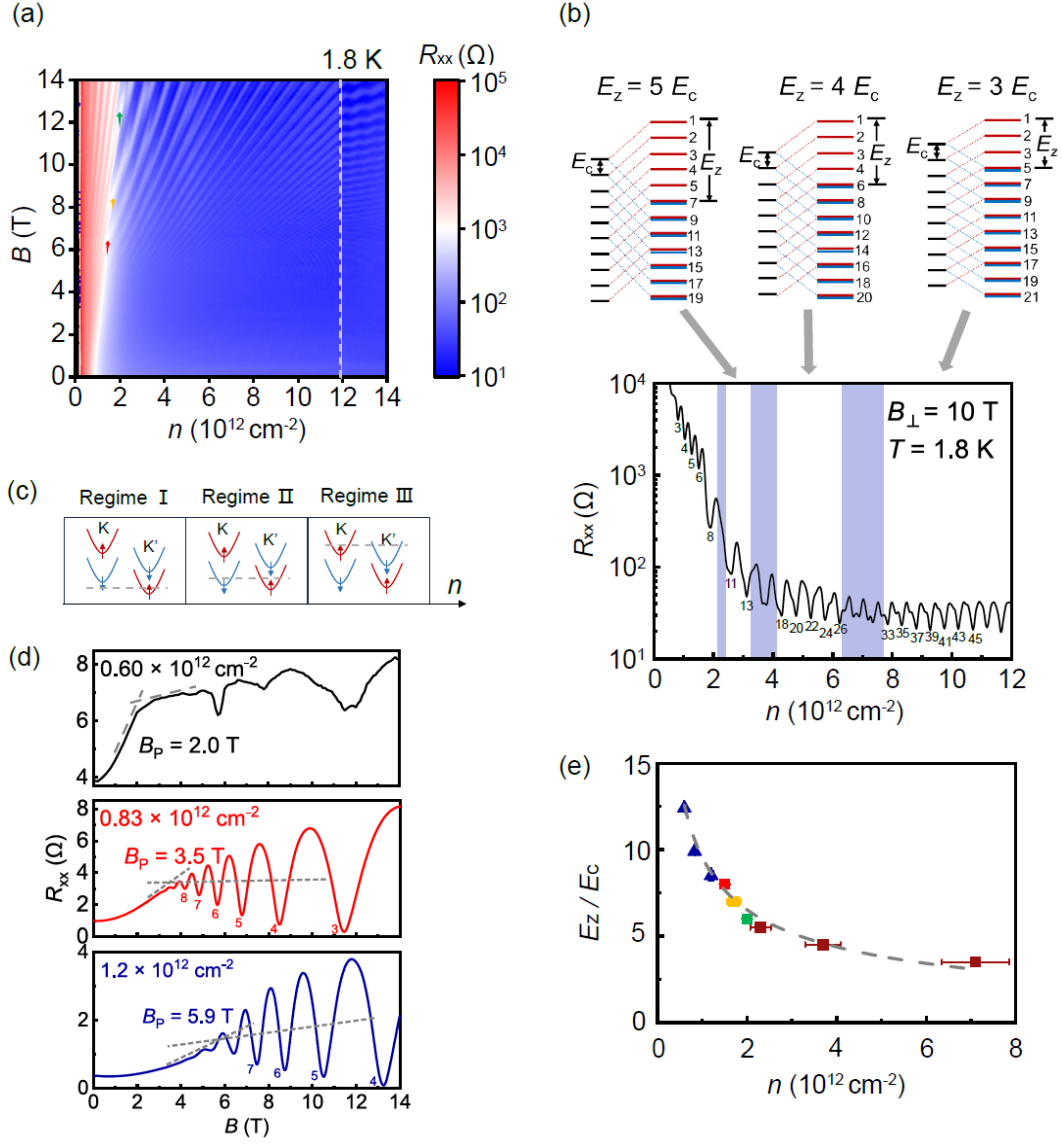

**Figure 3.** (a) $R_{xx}$ as a function of $n$ and $B$ measured at $T = 1.8$ K. The second spin-split band starts to be populated at $n \sim 12.5 \times 10^{12}$ cm$^{-2}$, indicated by the dashed line. (b) Measured $R_{xx}$ as a function of $n$ at $B = 10$ T and $T = 1.8$ K. At low electron densities, SdH oscillations are single-fold degenerate. At larger electron densities, SdH oscillations are two-fold degenerate, with the $\nu_{LL}$ sequences alternating between odd- and even-integer dominance, demonstrating the density-dependent nature of spin susceptibility. Blue-shaded regions highlight the density ranges of even-odd transitions. Schematic Landau level (LL) structures of the MoSe$_2$ conduction band at different $E_z/E_c$ ratios are shown in (b). (c) Schematic illustration of the conduction band minima at the K and K' points of monolayer MoSe$_2$ under finite magnetic fields and with different Fermi energies. (d) $R_{xx}$ versus $B$ measured between $n = 0.6 \times 10^{12}$ cm$^{-2}$ to $1.2 \times 10^{12}$ cm$^{-2}$ at $T = 1.8$ K, illustrating the evolution of magnetoresistance with $n$. (e) Ratio of $E_z/E_c$ plotted as a function of $n$.

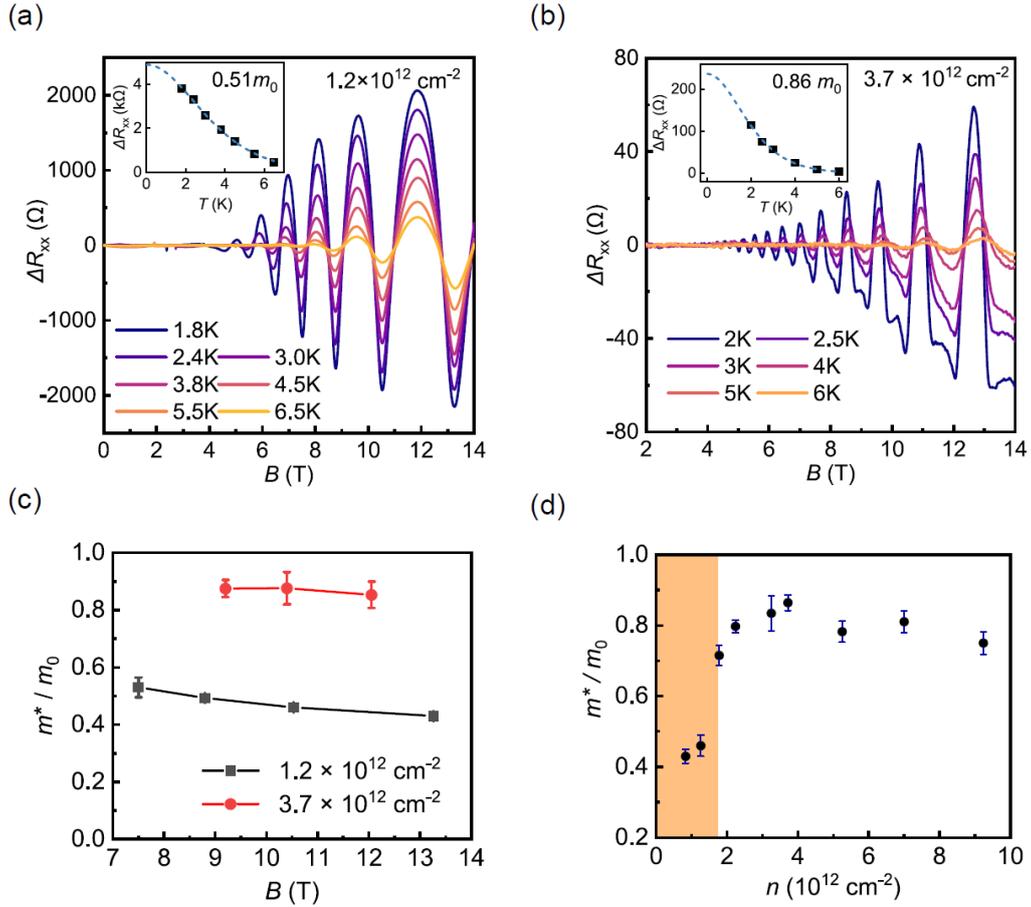

**Figure 4.** Temperature dependence of the SdH oscillation amplitude at $n = 1.2 \times 10^{12}$ cm$^{-2}$ (a) and $n = 3.7 \times 10^{12}$ cm$^{-2}$ (b). Insect: Effective mass $m^* = 0.51 m_0$ (a) and $m^* = 0.86 m_0$ (b) can be extracted by fitting with Lifshitz-Kosevitch formula. (c) Magnetic field dependence of $m^*$ at $n = 1.2 \times 10^{12}$ cm$^{-2}$ (red dot) and $n = 3.7 \times 10^{12}$ cm$^{-2}$ (black square). (d) $m^*$ as a function of $n$, which exhibits a sudden drop when the system becomes spin and valley polarized (orange-shaded region).